# Graphene-Based Artificial Dendrites for Bio-Inspired Learning in Spiking Neuromorphic Systems

*Samuel Liu[1,2], Deji Akinwande[1,2], Dmitry Kireev[1,2,3], and Jean Anne C. Incorvia[1,2]*

[1] Department of Electrical and Computer Engineering, The University of Texas at Austin, Austin, Texas 78712, USA

[2] Microelectronics Research Center, The University of Texas at Austin, Austin, Texas 78758, USA

[3] Department of Biomedical Engineering, University of Massachusetts Amherst, Amherst, Massachusetts 01003, USA

ABSTRACT (Word Style "BD_Abstract"). Analog neuromorphic computing systems emulate the parallelism and connectivity of the human brain, promising greater expressivity and energy efficiency compared to digital systems. Though many devices have emerged as candidates for artificial neurons and artificial synapses, there have been few device candidates for artificial dendrites. In this work, we report on biocompatible graphene-based artificial dendrites (GrADs) that can implement dendritic processing. By using a dual side-gate configuration, current applied through a Nafion membrane can be used to control device conductance across a trilayer graphene channel, showing spatiotemporal responses of leaky recurrent, alpha, and gaussian dendritic potentials. The devices can be variably connected to enable higher order neuronal responses, and we show through data-driven spiking neural network classification simulations that overall spiking activity is reduced by up to 15% without accuracy loss while low frequency operation is stabilized. This positions the GrADs as strong candidates for energy efficient bio-interfaced spiking neural networks.

TEXT (Word Style "TA_Main_Text").

Artificial neural networks (ANNs) are rapidly proliferating as a useful tool to process large amounts of data that cannot easily be analyzed with other methods[1,2]. However, due to the separation of compute and memory in von Neumann architecture, the speed and energy demands of state-of-the-art neural network models require new devices and architectures for further development[3,4]. Neuromorphic computing draws inspiration from the brain, combining artificial neurons and synapses to implement ANNs in hardware[5–8]. In particular, spiking neural networks (SNNs) process information in time and location-dependent spikes for bio-mimetic low-energy computation[6]. The requirements for ANN and SNN devices depend on their intended use, *e.g.*, CMOS-compatible devices for use with state-of-the-art electronics and biocompatible devices for use in health and bio-interfaces. Many devices are proposed to build these systems[9–25], but only a subset meets the performance requirements[26–29], few can realize the bio-compatibility necessary to interface with biological systems[30–33], and fewer leverage advanced biological behavior such as that of dendrites[34–36]. In biological systems, dendrites branch out from the neuronal body and process incoming spikes into non-spiking spatiotemporal signals. For example, the commonly known leaky integrate-and-fire model of neuronal behavior is composed of one leaky recurrent dendrite and a soma for an activation threshold. However, in many biological systems, each neuron can have multiple sets of dendrites independently processing incoming spikes, allowing for a wide variety of neuronal configurations[37]. Artificial dendrite function is an underexplored topic in neuromorphic computing that is considered biologically and computationally important for expressive SNNs. Previous work on analog neuromorphic devices for artificial dendrites has focused on using CMOS circuitry to emulate dendritic functions, requiring circuit overhead when designing a neuronal configuration with multiple dendrites[36,38]. Another work has proposed the use of the time-independent behavior of a volatile memristor to eliminate low amplitude noise in incoming signals and has also presented simulation results showing that dendrite-like behavior provides performance enhancement in non-spiking neural networks[39]. An artificial dendrite device candidate tailored for SNNs could greatly enhance the functionality of analog neuromorphic systems.

In this work, we emulate spatiotemporal dendrite dynamics on a chip by designing and measuring a graphene-based artificial dendrite (GrAD) for use in bio-compatible neuromorphic computing, and we show the device can aid in processing low spiking activity data. We have previously shown that transistors formed using bio-compatible materials, *i.e.*, graphene as the channel material and Nafion-117 in place of a gate insulator, can be operated as artificial synapses with plasticity that implements weight normalization techniques for online learning[40]. Here, by designing dual-gate operation of the graphene devices, we measure that fabricated macroscale GrADs (mGrADs) and microscale GrADs (μGrADs) emulate three different dendritic spatiotemporal signals: leaky recurrent, alpha, and gaussian. We then show combined operation of two mGrADs results in higher complexity time-dependent dynamics. We then show through data-driven simulations that dendrites can enhance SNNs by increasing stability when training at low energy, and that they reduce energy dissipation by lowering overall spiking activity by 15% without accuracy loss.

An example image of a multipolar neuron is shown in the left panel of Fig. 1a. The goal is to emulate dendritic behavior in a single dendritic branch, shown by the dotted-line box. A diagram of the GrAD is shown in the Fig. 1a middle panel. Trilayer graphene functions as a channel material between source (S) and drain (D) contacts, and Nafion-117 is used as a gate insulator. Two gates denoted as the input gate (InG) and tuning gate (TG) are placed to the sides of the channel underneath the Nafion, on the same layer as the S/D contacts, to provide a side-gating effect. Current signals $I_{InG}$ and $I_{TG}$ control the conductance of the channel, which is read by applying a read voltage $V_{read}$ between the D and S terminals with the S grounded. For this work, two types of GrAD devices were fabricated: macroscale (mGrAD, ~50 mm$^2$) and microscale (μGrAD, 40 μm × 40 μm). The right panel of Fig. 1a depicts the circuit schematic symbol to be used in subsequent figures. Figure 1b shows the transfer curve of a μGrAD operated as an electrolytic transistor, where gate voltage is forward and reverse swept twice between -1 V and 1 V at a ramp rate of 50 mV/sec across terminals InG and S, with a fixed drain voltage of 0.1 V. Drain current and gate current are recorded. The recorded charge neutrality point of the μGrAD is approximately 0.5 V, consistent with p-doped graphene transistors fabricated using wet transfer. The characteristics of the transfer curve are consistent with results seen in electrolytic transistors. We see hysteresis of the transfer characteristics for both drain current and gate current, indicating that there is a memory effect on the timescale of the ramp rate, consistent with the time constant of ionic movement. Due to the size of the device and Nafion gating material, there is high leakage current observed, reaching 10 nA at applied voltage of 1.0 V. When operating the GrAD using current operation through the InG with a floating TG, the device shows artificial synapse characteristics, shown in Fig. 1c, where positive and negative pulse trains are applied for several cycles showing distinct synaptic weight levels. This long-term potentiation was analyzed in our previous work, facilitated by the shifting of ionic distribution within the dry Nafion-117 membrane[40].

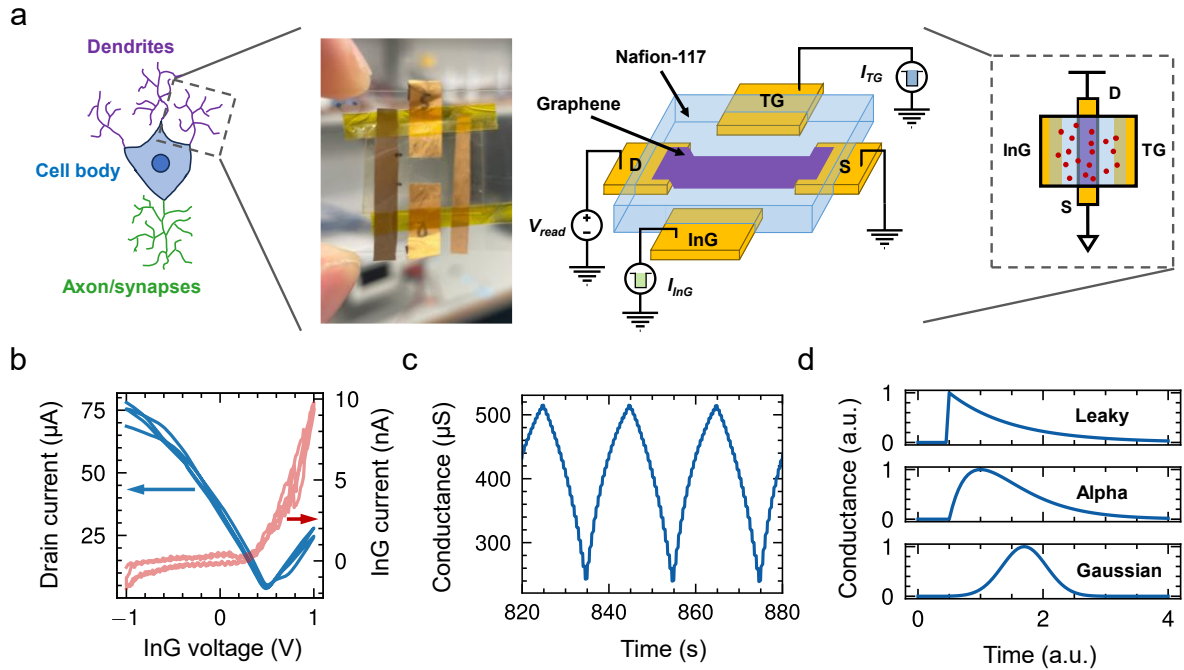

**Figure 1. GrAD design and single gate operation. a)** (Left) Diagram of neuron depicting three dendritic branches (purple), each of which is implemented by a GrAD. (Middle) Picture of mGrAD along with diagram of the GrAD structure, where the dynamics are controlled using InG and TG and readout is through voltage applied across the graphene channel (purple) through terminals D and S. Nafion-117 gating material is shown in light blue, and Au contacts are shown in gold. (Right) Circuit schematic symbol of the GrAD. **b)** Transfer characteristics of a μGrAD with two forward and backward sweeps of -1 V to 1 V. D-S current is shown in blue, and InG-S current shown in red. **c)** Single gate current operation of a μGrAD showing synaptic characteristics. **d)** Leaky recurrent, alpha, and gaussian dendritic kernel waveforms commonly recorded in biological dendrites.

Three time-dependent waveforms, the leaky recurrent, alpha, and gaussian kernels, are chosen from a set identified in Ref. [34] as target dendrite behaviors, shown in Fig. 1d. To introduce a controlled inhibition of the dendritic conductance, a constant positive bias current is applied to the TG, shown in Fig. 2a. Then, to stimulate the leaky recurrent waveform, the spiking signals entering the dendrite are represented as square wave current pulses applied to the InG. In Fig. 2b, $I_{TG}$ is 100 nA, and $I_{InG}$ is 1 ms long, -10 μA amplitude, applied to a mGrAD at input frequencies of 1 Hz, 2 Hz, and 5 Hz. The conductance of the graphene channel is tracked to monitor the state of the device. A threshold conductance is set at 1 mS to emulate the effect of a soma. It is observed that the GrAD conductance integrates $I_{InG}$ is on and leaks back to lower conductance when $I_{InG}$ is off. The combination of a single leaky recurrent dendrite and a soma models the membrane potential dynamics of a leaky integrate-and-fire (LIF) neuron. By taking the inverse of the time taken to reach the threshold from rest, the output spiking frequency can be inferred. In Fig. 2b, the times are labeled $t_1$, $t_2$, and $t_3$. The frequency-based activation function of a LIF neuron model is in the form of a Rectified Linear Unit (ReLU), where the threshold for rectification and slope of the ReLU function are mediated by the strength of leaking. This is reflected in the mGrAD, where the output spiking frequency is further inhibited by an increase in current applied at the TG, shown in Fig. 2c. This shows that the GrAD can not only have LIF artificial neuron function, but that the response is tunable by a DC current applied at the TG.

To emulate the other two dendrite kernels, alpha and gaussian, the input spikes are represented as triangular wave current pulses applied at the InG, and the inhibiting constant current is applied at the TG like the previous case, depicted in Fig. 2d. Figure 2e shows the resulting waveform for $I_{InG}$ amplitudes of -1 μA to -4 μA with a half-max pulse duration of 500 ms, and $I_{TG}$ = 200 nA. The GrAD response is qualitatively similar to the alpha kernel. The characteristic equation of the alpha kernel is as follows:

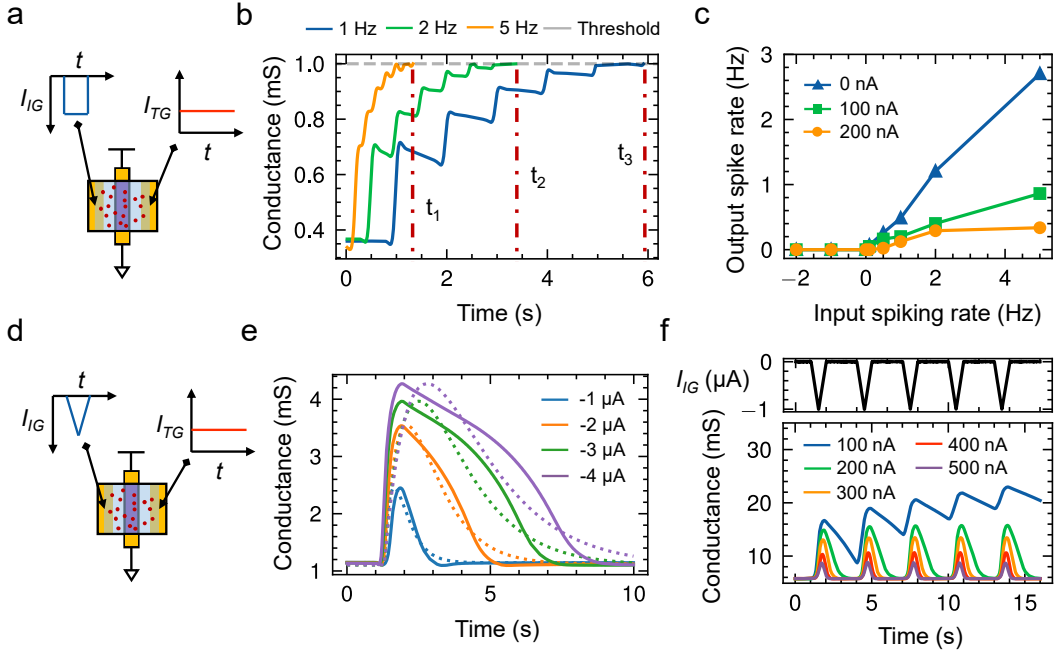

**Figure 2. mGrAD time-dependent response. a)** A square current pulse applied to the InG and constant current applied at the TG generates the leaky recurrent response. **b)** Leaky recurrent mGrAD device response to the InG pulse trains of varying frequency (blue for 1 Hz, green for 2 Hz, and orange for 5 Hz). Time taken to reach threshold is denoted with the red dashed line. **c)** Activation function of leaky recurrent dendrite with varying DC current applied to the TG (blue triangles for 0 nA, green squares for 100 nA, and orange circles for 200 nA). **d)** A triangular current pulse applied to the InG and constant current applied at the TG generates the alpha response. **e)** Alpha mGrAD device response for varying $I_{InG}$ amplitude (solid lines) compared to an ideal alpha kernel response (dotted lines). **f)** Alpha and gaussian waveform generation at varying $I_{TG}$ (blue for 100 nA, green for 200 nA, orange for 300 nA, red for 400 nA, and purple for 500 nA).

$$g_{alpha} = \frac{g_{max}t}{\tau}\exp\left(-\frac{t-\tau}{\tau}\right) \quad\quad\quad (1)$$

where $t$ is time, $\tau$ is the time constant, and $g_{max}$ is the peak conductance reached. To find the fit of the equation, the minimum conductance was subtracted from the experimental data and the initial delay of 1 s before the waveform arrives is eliminated. The best fit for $\tau$ was found to be 0.35 sec, 0.82 sec, 1.24 sec, and 1.51 sec for -1 µA, -2 µA, -3 µA, and -4 µA, respectively, shown as dotted lines. There is also an additional delay of 0.3 sec between the start of the triangular pulse applied to the InG and the beginning of the alpha kernel waveform resulting from the device. It is observed that as $I_{InG}$ becomes larger, the deviation from an ideal alpha kernel becomes larger as well. In particular, the location of the conductance peak and a hump during the decay are the places with the largest deviation. However, when limited to small values, the waveforms match the alpha kernel.

The mGrAD can be further tuned to show a gaussian kernel by applying different $I_{TG}$. In Fig. 2f, a series of 5 triangular pulses at the IG were applied with an amplitude of -1 µA and half-max duration of 500 ms, shown in black. TG bias currents from 100 nA to 500 nA were applied, and the conductance of the channel was tracked. As the bias current applied to the TG is increased, both $\tau$ and $g_{max}$ decrease. Between 300 nA and 500 nA of bias, the output of the GrAD was measured to be symmetric in time, matching the shape of a gaussian kernel over that of an alpha kernel. The characteristic equation of the gaussian kernel can be described as follows:

$$g_{gauss} = g_{max} \exp\left(\frac{(t-t_0)^2}{2\tau_{gauss}^2}\right) \qquad (2)$$

where $\tau_{gauss}$ is the time constant associated with the width of the gaussian kernel and $t_0$ is the time at which the center of the gaussian kernel occurs. The time constant was found to be 0.85 sec, 0.67 sec, and 0.44 sec for 300 nA, 400 nA, and 500 nA of bias, respectively. The amplitude of conductance change $g_{max}$ also changed with different $I_{TG}$, resulting in 7.8 µS, 4.7 µS, and 2.7 µS for 300 nA, 400 nA, and 500 nA respectively.

While mGrAD shows potentially useful emulation of dendritic kernels, the ability to combine these signals to form more complex time-dependent dynamics for neurons is necessary for larger scale implementation. Figure 3a depicts how the GrAD devices can be connected to form a full dendritic unit. A demonstration of this dendritic unit was implemented using two GrAD devices connected in a circuit shown in Fig. 3b. Two different AC signals are respectively applied to the input gates, while a bias DC current $I_{TG} = 200$ nA was applied to both TGs. Figure 3c shows the two input signals in purple and red in the top panel. The resulting waveform shown in the bottom panel is the combined output current of the two mGrADs that represents the effective membrane potential change over time of the dendritic unit. The results demonstrate that multiple GrADs can be flexibly connected to implement the desired dendritic unit configuration.

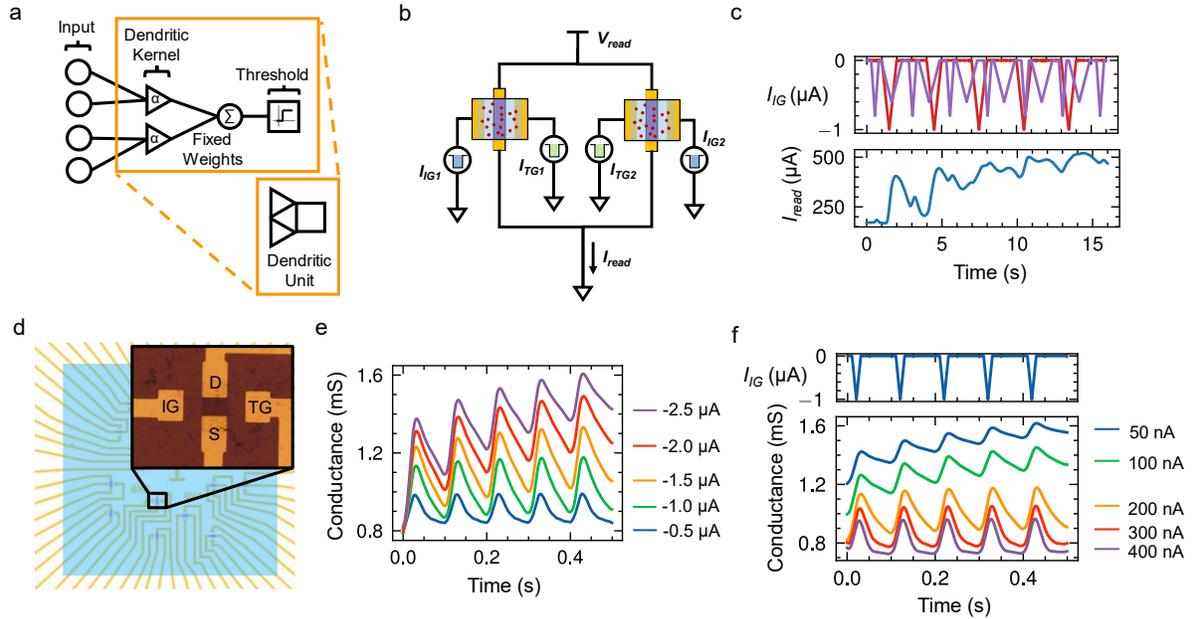

**Figure 3. Multiple GrAD and μGrAD time-dependent response. a)** Symbolic representation of dendritic unit. **b)** Circuit diagram of dendritic unit. **c)** Experimental output of dendritic unit consisting of two dendrites. **d)** Schematic of μGrAD dual side gate device layout. Contacts are shown in pink and graphene channel is shown in purple. **e)** Alpha waveform generation at varying $I_{IG}$. **g)** Alpha and gaussian waveform generation at varying $I_{TG}$.

To demonstrate the potential for large scale implementation of dendritic devices, μGrADs were fabricated by drop casting Nafion-117 on top of graphene transistors with two side gates, shown in Fig. 3d. The same dendrite characterization done previously for the mGrADs was applied to the μGrADs. Figure 3e shows the output waveform for repeated triangular pulses applied to the InG with varying amplitude between -0.5 μA and -2.5 μA with a half-max pulse duration of 10 ms, with $I_{TG}$ = 200 nA. A low pass Butterworth filter with a cutoff frequency of 25 Hz was used to eliminate noise introduced by the measurement setup. The measured data from the μGrADs shows similar behavior to the mGrAD device, where the change in conductance was larger as the pulse amplitude increased. While operation of the μGrAD was within a reduced conductance ratio from 0.8 mS to 1.6 mS compared to the conductance ratio of between l mS to 4 mS in the mGrAD, the time constant $\tau$ of the fit applied from Eqn. 1 was more uniform between different amplitudes, at 14.2 ms, 15.1 ms, 18.9 ms, 22.4 ms, and 25.4 ms respectively for -0.5 μA, -1.0 μA, -1.5 μA, -2.0 μA and -2.5 μA, showing more similarly shaped responses. Additionally, with the chosen operational range, smaller amplitude pulses with smaller pulse widths were required to change the conductance of the device, indicating that scaling is favorable for increasing the speed of operation as well as lowering the current requirements. Figure 3f shows that the current applied at the TG can also be used to tune the time constant of dendritic response. While $I_{TG}$ for the μGrAD is similar in magnitude to that applied to the mGrAD, the results indicate that faster operation of the device is enabled due to microscaling.

After characterizing the mGrADs and μGrADs, the experimental time-dependent conductance response was used to simulate supervised learning in a spiking neural network (SNN). The Fashion-MNIST clothing article classification task was chosen as the benchmark. Training is simulated using custom modules in the Norse framework[41] and a toolset based on PyTorch[42] that can model behaviors of individual neurons within a network. The network architecture used for this simulation is a modified multilayer perceptron with one hidden layer of 200 units, shown in Fig. 4a. The neurons in the hidden layer and output layer are replaced with dendritic units, where the number of dendrites per unit can be later specified. Additionally, a modification to connections between neurons is implemented when there are multiple dendrites in a dendritic unit, following an architecture described in Ref. [39]. The connections of the network emulate the variable connectivity of multipolar neurons with multiple dendritic branches[43]. If the dendritic units in the hidden layer have X dendrites, then the neurons of the input layer are split into X groups. Each group is then connected to their corresponding dendrite *e.g.*, neurons of the first group are connected to the first dendrite in each unit. The notation XH:YO describes the network configuration, where X and Y are the number of dendrites per dendritic unit in the hidden layer and output layer, respectively. In addition to the components shown in Fig. 4a, a Poisson encoder converts the pixel brightness values of each image into triangular spike trains, and a softmax is applied to the output potentials to calculate the gradient. The inset of Fig. 4a shows the analog crossbar implementation of the network, where the synapses are the weighted connections emulated using μBLAST devices[40]. The summed currents along the columns (green) then feed into the InG of a GrAD, where the conductance of the device determines the firing of the soma. To evaluate the impact of introducing dendrites to a SNN, a baseline network consisting of ideal LIF neurons described in Supplementary Note 1 was evaluated alongside the dendritic network configurations.

In order to characterize the online learning performance of the dendritic network under varying operating conditions, sampling time per image and maximum input frequency are swept. In Fig. 4b-c, the maximum accuracy after 20 epochs of two configurations of dendritic networks, 2H:2O and 4H:2O, are compared to the LIF network for varying sampling time per image. The maximum input spiking frequency is held fixed at 50 Hz. The maximum accuracy of the network is consistent between the different configurations. However, the LIF network and the 2H:2O network both have a drop-off point where the network does not train below a certain sampling time per image, dropping to 10% accuracy, the equivalent of a random guess. This occurs at 100 ms for the LIF network and 50 ms for the 2H:2O network. The 4H:2O network can be trained for the full range of sampling time per image, indicating that performance for an optimal configuration of dendritic network can have enhanced stability even with shorter sampling time. Additionally, in Fig. 4c, the spike number characterizes the energy efficiency of the network since the spiking activity within the network is directly correlated to energy dissipation. For sampling time per image where the networks are successfully trained, the 4H:2O and 2H:2O network have respectively 15% and 9% average reduced spiking activity compared to the LIF network, indicating that increased energy efficiency is an additional benefit of a dendritic network. A sweep of maximum input frequency in Fig. 4d-e where sampling time per image is held constant at 200 ms yields similar observations. The LIF network fails to train when maximum input frequency drops below 40 Hz in contrast to the dendritic network, which successfully trained for all input frequencies evaluated. For the dendritic network, spiking activity for frequencies that resulted in successful training was again 15% lower than the spiking activity for the LIF network, corroborating the previous result.

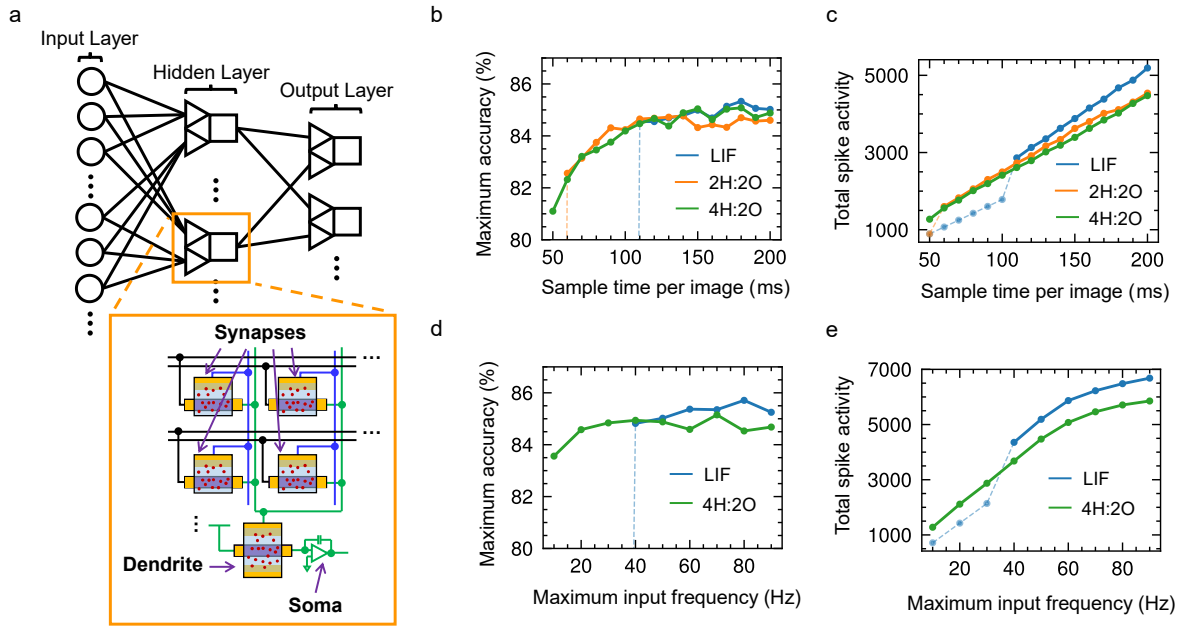

**Figure 4. Dendritic spiking neural network simulations. a)** Symbolic representation of dendritic multilayer perceptron. Circuit representation of analog crossbar is inset. **b)** Maximum accuracy after training for 20 epochs as a function of sample time per image. The drop-off settles at accuracy of 10% below indicated point. **c)** Total spike activity of network as a function of sample time per image. Faded coloring indicates accuracy of 10% for corresponding sample time per image. **d)** Maximum accuracy after training for 20 epochs as a function of maximum input frequency. **e)** Total spike activity of network as a function of maximum input frequency.

In conclusion, we have demonstrated dendritic behavior in graphene/Nafion devices with a dual-gate structure at timescales compatible with biological signals. We have shown that there is a high degree of tunability of the time-dependent conductance signal, which can be used to represent how a dendrite responds in time to a train of spiking input signals. We show that microdevice scaling can increase the speed of operation and reduce current requirements. Through neural network simulations driven by this experimental behavior, we show that SNNs can benefit greatly from dendritic configuration, showing significantly higher training stability at low power operation and lower spiking activity. These characteristics make the GrAD a promising building block for networks bridging bioelectronics and neuromorphic computing.

METHODS

mGrAD Device Fabrication

The graphene was prepared by first mounting it on tattoo paper for transfer. Monolayer CVD grown large-scale graphene on copper (Grolltex) was taped onto a silicon wafer and spin-coated with PMMA at 2500 rpm for 60 s, resulting in a ~200 nm thick layer. The sampled is then hard-baked at 200 degrees C for 15-20 min. The copper is then etched away by placing the sample in 0.1 M ammonium persulfate. The PMMA/graphene film is then wet transferred onto temporary tattoo paper. The gate contacts are prepared by evaporating gold onto an EVA/PET film. The two gates are then contacted to a Nafion-117 pre-processed film and heated at 150 degrees C to adhere. Adhesive gold contacts are then attached on top of the Nafion-117 film to form the source and drain contacts and prepare for graphene transfer. The graphene tattoo paper is then soaked in DI water and transferred once the edges start delaminating, forming a channel across the Nafion.

## μGrAD Device Fabrication

CVD grown graphene is prepared by spin-coating PMMA and etching away copper in the method described in the previous section. The graphene was then transferred onto Au/Cr (90/10 nm thick) contacts patterned on an Si/SiO$_2$ wafer. The graphene channel is then patterned and excess graphene is etched away using oxygen plasma. Photostructurable polyimide HD8820 was used in the last step to form the passivation. The devices are diced from the wafer and drop coated with Nafion-117 solution (Sigma-Aldrich) and hard-baked at 150 degrees.

## Device Measurement Setup

The devices were measured using two high-precision source/measure unit Agilent 2902B. One of the SMUs is used to apply 0.1 V of the drain-source potential, while the gate is used in the current-pulsing mode with various pulse shape to induce conductance changes. A third SMU is used to apply a small constant positive current to tune the dendritic behavior.

ASSOCIATED CONTENT

(Word Style "TE_Supporting_Information"). **Supporting Information**. A listing of the contents of each file supplied as Supporting Information should be included. For instructions on what should be included in the Supporting Information as well as how to prepare this material for publications, refer to the journal's Instructions for Authors.

The following files are available free of charge.

brief description (file type, i.e., PDF)

brief description (file type, i.e., PDF)

AUTHOR INFORMATION

**Corresponding Author**

*(Word Style "FA_Corresponding_Author_Footnote"). * (Word Style "FA_Corresponding_Author_Footnote"). Give contact information for the author(s) to whom correspondence should be addressed.

**Author Contributions**

Temp

**Funding Sources**

This research is partially supported by the National Science Foundation Graduate Research Fellowship under Grant No. 2021311125 and partially supported by the National Science Foundation under CCF award 2246855.

ACKNOWLEDGMENT

Temp

REFERENCES

(1)     Hopfield, J. J. Artificial Neural Networks. *IEEE Circuits and Devices Magazine* **1988**, *4* (5), 3–10. https://doi.org/10.1109/101.8118.

(2)     Krogh, A. What Are Artificial Neural Networks? *Nat Biotechnol* **2008**, *26* (2), 195–197. https://doi.org/10.1038/nbt1386.

(3)     Wulf, Wm. A.; McKee, S. A. Hitting the Memory Wall. *ACM SIGARCH Computer Architecture News* **1995**, *23* (1), 20–24. https://doi.org/10.1145/216585.216588.

(4)     Angizi, S.; He, Z.; Reis, D.; Hu, X. S.; Tsai, W.; Lin, S. J.; Fan, D. Accelerating Deep Neural Networks in Processing-in-Memory Platforms: Analog or Digital Approach? In *2019 IEEE Computer Society Annual Symposium on VLSI (ISVLSI)*; IEEE, 2019; pp 197–202. https://doi.org/10.1109/ISVLSI.2019.00044.

(5)     Schuman, C. D.; Kulkarni, S. R.; Parsa, M.; Mitchell, J. P.; Date, P.; Kay, B. Opportunities for Neuromorphic Computing Algorithms and Applications. *Nat Comput Sci* **2022**, *2* (1), 10–19. https://doi.org/10.1038/s43588-021-00184-y.

(6)     Roy, K.; Jaiswal, A.; Panda, P. Towards Spike-Based Machine Intelligence with Neuromorphic Computing. *Nature* **2019**, *575* (7784), 607–617. https://doi.org/10.1038/s41586-019-1677-2.

(7)     Marković, D.; Mizrahi, A.; Querlioz, D.; Grollier, J. Physics for Neuromorphic Computing. *Nature Reviews Physics* **2020**, *2* (9), 499–510. https://doi.org/10.1038/s42254-020-0208-2.

(8)     Burr, G. W.; Shelby, R. M.; Sebastian, A.; Kim, S.; Kim, S.; Sidler, S.; Virwani, K.; Ishii, M.; Narayanan, P.; Fumarola, A.; Sanches, L. L.; Boybat, I.; Le Gallo, M.; Moon, K.; Woo, J.; Hwang, H.; Leblebici, Y. Neuromorphic Computing Using Non-Volatile Memory. *Adv Phys X* **2017**, *2* (1), 89–124. https://doi.org/10.1080/23746149.2016.1259585.

(9)     La Barbera, S.; Ly, D. R. B.; Navarro, G.; Castellani, N.; Cueto, O.; Bourgeois, G.; De Salvo, B.; Nowak, E.; Querlioz, D.; Vianello, E. Narrow Heater Bottom Electrode-Based Phase Change Memory as a Bidirectional Artificial Synapse. *Adv Electron Mater* **2018**, *4* (9). https://doi.org/10.1002/aelm.201800223.

(10)    Wong, H. S. P.; Raoux, S.; Kim, S.; Liang, J.; Reifenberg, J. P.; Rajendran, B.; Asheghi, M.; Goodson, K. E. Phase Change Memory. *Proceedings of the IEEE* **2010**, *98* (12), 2201–2227. https://doi.org/10.1109/JPROC.2010.2070050.

(11)    Kumar, S.; Wang, X.; Strachan, J. P.; Yang, Y.; Lu, W. D. Dynamical Memristors for Higher-Complexity Neuromorphic Computing. *Nat Rev Mater* **2022**, *7* (7), 575–591. https://doi.org/10.1038/s41578-022-00434-z.

(12)    Wang, M.; Cai, S.; Pan, C.; Wang, C.; Lian, X.; Zhuo, Y.; Xu, K.; Cao, T.; Pan, X.; Wang, B.; Liang, S.-J.; Yang, J. J.; Wang, P.; Miao, F. Robust Memristors Based on Layered Two-Dimensional Materials. *Nat Electron* **2018**, *1* (2), 130–136. https://doi.org/10.1038/s41928-018-0021-4.

(13) Milo, V.; Malavena, G.; Monzio Compagnoni, C.; Ielmini, D. Memristive and CMOS Devices for Neuromorphic Computing. *Materials* **2020**, *13* (1), 166. https://doi.org/10.3390/ma13010166.

(14) Zhou, G.; Wang, Z.; Sun, B.; Zhou, F.; Sun, L.; Zhao, H.; Hu, X.; Peng, X.; Yan, J.; Wang, H.; Wang, W.; Li, J.; Yan, B.; Kuang, D.; Wang, Y.; Wang, L.; Duan, S. Volatile and Nonvolatile Memristive Devices for Neuromorphic Computing. *Adv Electron Mater* **2022**, 2101127. https://doi.org/10.1002/aelm.202101127.

(15) Siddiqui, S. A.; Dutta, S.; Tang, A.; Liu, L.; Ross, C. A.; Baldo, M. A. Magnetic Domain Wall Based Synaptic and Activation Function Generator for Neuromorphic Accelerators. *Nano Lett* **2020**, *20* (2), 1033–1040. https://doi.org/10.1021/acs.nanolett.9b04200.

(16) Grollier, J.; Querlioz, D.; Camsari, K. Y.; Everschor-Sitte, K.; Fukami, S.; Stiles, M. D. Neuromorphic Spintronics. *Nat Electron* **2020**, *3* (7), 360–370. https://doi.org/10.1038/s41928-019-0360-9.

(17) Jung, S.; Lee, H.; Myung, S.; Kim, H.; Yoon, S. K.; Kwon, S.-W.; Ju, Y.; Kim, M.; Yi, W.; Han, S.; Kwon, B.; Seo, B.; Lee, K.; Koh, G.-H.; Lee, K.; Song, Y.; Choi, C.; Ham, D.; Kim, S. J. A Crossbar Array of Magnetoresistive Memory Devices for In-Memory Computing. *Nature* **2022**, *601* (7892), 211–216. https://doi.org/10.1038/s41586-021-04196-6.

(18) Leonard, T.; Liu, S.; Jin, H.; Incorvia, J. A. C. Stochastic Domain Wall-Magnetic Tunnel Junction Artificial Neurons for Noise-Resilient Spiking Neural Networks. *Appl Phys Lett* **2023**, *122* (26). https://doi.org/10.1063/5.0152211.

(19) Liu, S.; Xiao, T. P.; Kwon, J.; Debusschere, B. J.; Agarwal, S.; Incorvia, J. A. C.; Bennett, C. H. Bayesian Neural Networks Using Magnetic Tunnel Junction-Based Probabilistic in-Memory Computing. *Frontiers in Nanotechnology* **2022**, *4*. https://doi.org/10.3389/fnano.2022.1021943.

(20) Leonard, T.; Liu, S.; Alamdar, M.; Jin, H.; Cui, C.; Akinola, O. G.; Xue, L.; Xiao, T. P.; Friedman, J. S.; Marinella, M. J.; Bennett, C. H.; Incorvia, J. A. C. Shape-Dependent Multi-Weight Magnetic Artificial Synapses for Neuromorphic Computing. *Adv Electron Mater* **2022**, *8* (12), 2200563. https://doi.org/10.1002/aelm.202200563.

(21) Hu, X.; Cui, C.; Liu, S.; Garcia-Sanchez, F.; Brigner, W. H.; Walker, B. W.; Edwards, A. J.; Xiao, T. P.; Bennett, C.; Hassan, N.; Frank, M. P.; Incorvia, J. A. C.; Friedman, J. Magnetic Skyrmions and Domain Walls for Logical and Neuromorphic Computing. *Neuromorphic Computing and Engineering* **2023**. https://doi.org/10.1088/2634-4386/acc6e8.

(22) Gkoupidenis, P.; Schaefer, N.; Garlan, B.; Malliaras, G. G. Neuromorphic Functions in PEDOT:PSS Organic Electrochemical Transistors. *Advanced Materials* **2015**, *27* (44), 7176–7180. https://doi.org/10.1002/adma.201503674.

(23) Melianas, A.; Quill, T. J.; LeCroy, G.; Tuchman, Y.; Loo, H. v.; Keene, S. T.; Giovannitti, A.; Lee, H. R.; Maria, I. P.; McCulloch, I.; Salleo, A. Temperature-Resilient Solid-State

Organic Artificial Synapses for Neuromorphic Computing. *Sci Adv* **2020**, *6* (27). https://doi.org/10.1126/sciadv.abb2958.

(24)  van de Burgt, Y.; Lubberman, E.; Fuller, E. J.; Keene, S. T.; Faria, G. C.; Agarwal, S.; Marinella, M. J.; Alec Talin, A.; Salleo, A. A Non-Volatile Organic Electrochemical Device as a Low-Voltage Artificial Synapse for Neuromorphic Computing. *Nat Mater* **2017**, *16* (4), 414–418. https://doi.org/10.1038/nmat4856.

(25)  Huang, H.; Yang, R.; Tan, Z.; He, H.; Zhou, W.; Xiong, J.; Guo, X. Quasi-Hodgkin–Huxley Neurons with Leaky Integrate-and-Fire Functions Physically Realized with Memristive Devices. *Advanced Materials* **2019**, *31* (3), 1803849. https://doi.org/10.1002/adma.201803849.

(26)  Sun, X.; Yu, S. Impact of Non-Ideal Characteristics of Resistive Synaptic Devices on Implementing Convolutional Neural Networks. *IEEE J Emerg Sel Top Circuits Syst* **2019**, *9* (3), 570–579. https://doi.org/10.1109/JETCAS.2019.2933148.

(27)  Agarwal, S.; Plimpton, S. J.; Hughart, D. R.; Hsia, A. H.; Richter, I.; Cox, J. A.; James, C. D.; Marinella, M. J. Resistive Memory Device Requirements for a Neural Algorithm Accelerator. In *2016 International Joint Conference on Neural Networks (IJCNN)*; IEEE, 2016; pp 929–938. https://doi.org/10.1109/IJCNN.2016.7727298.

(28)  Xiao, T. P.; Bennett, C. H.; Feinberg, B.; Agarwal, S.; Marinella, M. J. Analog Architectures for Neural Network Acceleration Based on Non-Volatile Memory. *Appl Phys Rev* **2020**, *7* (3). https://doi.org/10.1063/1.5143815.

(29)  Xiao, T. P.; Feinberg, B.; Bennett, C. H.; Prabhakar, V.; Saxena, P.; Agrawal, V.; Agarwal, S.; Marinella, M. J. On the Accuracy of Analog Neural Network Inference Accelerators. *IEEE Circuits and Systems Magazine* **2022**, *22* (4), 26–48. https://doi.org/10.1109/MCAS.2022.3214409.

(30)  Keene, S. T.; Lubrano, C.; Kazemzadeh, S.; Melianas, A.; Tuchman, Y.; Polino, G.; Scognamiglio, P.; Cinà, L.; Salleo, A.; van de Burgt, Y.; Santoro, F. A Biohybrid Synapse with Neurotransmitter-Mediated Plasticity. *Nat Mater* **2020**, *19* (9), 969–973. https://doi.org/10.1038/s41563-020-0703-y.

(31)  Kim, Y.; Park, C. H.; An, J. S.; Choi, S.-H.; Kim, T. W. Biocompatible Artificial Synapses Based on a Zein Active Layer Obtained from Maize for Neuromorphic Computing. *Sci Rep* **2021**, *11* (1), 20633. https://doi.org/10.1038/s41598-021-00076-1.

(32)  Park, H.; Lee, Y.; Kim, N.; Seo, D.; Go, G.; Lee, T. Flexible Neuromorphic Electronics for Computing, Soft Robotics, and Neuroprosthetics. *Advanced Materials* **2020**, *32* (15). https://doi.org/10.1002/adma.201903558.

(33)  Robinson, D. A.; Foster, M. E.; Bennett, C. H.; Bhandarkar, A.; Webster, E. R.; Celebi, A.; Celebi, N.; Fuller, E. J.; Stavila, V.; Spataru, C. D.; Ashby, D. S.; Marinella, M. J.; Krishnakumar, R.; Allendorf, M. D.; Talin, A. A. Tunable Intervalence Charge Transfer in Ruthenium Prussian Blue Analog Enables Stable and Efficient Biocompatible Artificial Synapses. *Advanced Materials* **2023**, *35* (37). https://doi.org/10.1002/adma.202207595.

(34)    Tapson, J. C.; Cohen, G. K.; Afshar, S.; Stiefel, K. M.; Buskila, Y.; Wang, R. M.; Hamilton, T. J.; van Schaik, A. Synthesis of Neural Networks for Spatio-Temporal Spike Pattern Recognition and Processing. *Front Neurosci* **2013**, *7*. https://doi.org/10.3389/fnins.2013.00153.

(35)    Yang, S.; Gao, T.; Wang, J.; Deng, B.; Lansdell, B.; Linares-Barranco, B. Efficient Spike-Driven Learning With Dendritic Event-Based Processing. *Front Neurosci* **2021**, *15*. https://doi.org/10.3389/fnins.2021.601109.

(36)    Kaiser, J.; Billaudelle, S.; Müller, E.; Tetzlaff, C.; Schemmel, J.; Schmitt, S. Emulating Dendritic Computing Paradigms on Analog Neuromorphic Hardware. *Neuroscience* **2022**, *489*, 290–300. https://doi.org/10.1016/j.neuroscience.2021.08.013.

(37)    Stuart, G.; Spruston, N.; Häusser, M. *Dendrites*, 3rd ed.; Oxford University Press, 2016.

(38)    Cardwell, S. G.; Chance, F. S. Dendritic Computation for Neuromorphic Applications. In *Proceedings of the 2023 International Conference on Neuromorphic Systems*; ACM: New York, NY, USA, 2023; pp 1–5. https://doi.org/10.1145/3589737.3606001.

(39)    Li, X.; Tang, J.; Zhang, Q.; Gao, B.; Yang, J. J.; Song, S.; Wu, W.; Zhang, W.; Yao, P.; Deng, N.; Deng, L.; Xie, Y.; Qian, H.; Wu, H. Power-Efficient Neural Network with Artificial Dendrites. *Nat Nanotechnol* **2020**, *15* (9), 776–782. https://doi.org/10.1038/s41565-020-0722-5.

(40)    Kireev, D.; Liu, S.; Jin, H.; Patrick Xiao, T.; Bennett, C. H.; Akinwande, D.; Incorvia, J. A. C. Metaplastic and Energy-Efficient Biocompatible Graphene Artificial Synaptic Transistors for Enhanced Accuracy Neuromorphic Computing. *Nat Commun* **2022**, *13* (1), 4386. https://doi.org/10.1038/s41467-022-32078-6.

(41)    Pehle, C.; Pedersen, J. E. Norse - A Deep Learning Library for Spiking Neural Networks. Zenodo January 2021. https://doi.org/10.5281/zenodo.4422025.

(42)    Paszke, A.; Gross, S.; Massa, F.; Lerer, A.; Bradbury, J.; Chanan, G.; Killeen, T.; Lin, Z.; Gimelshein, N.; Antiga, L.; Desmaison, A.; Kopf, A.; Yang, E.; DeVito, Z.; Raison, M.; Tejani, A.; Chilamkurthy, S.; Steiner, B.; Fang, L.; Bai, J.; Chintala, S. PyTorch: An Imperative Style, High-Performance Deep Learning Library. In *Advances in Neural Information Processing Systems*; Wallach, H., Larochelle, H., Beygelzimer, A., d Alché-Buc, F., Fox, E., Garnett, R., Eds.; Curran Associates, Inc., 2019; Vol. 32.

(43)    Jossin, Y.; Cooper, J. A. Reelin, Rap1 and N-Cadherin Orient the Migration of Multipolar Neurons in the Developing Neocortex. *Nat Neurosci* **2011**, *14* (6), 697–703. https://doi.org/10.1038/nn.2816.